\documentclass[preprint,number,sort&compress,12pt]{elsarticle}

\usepackage{epsfig}

\usepackage{amssymb}

\usepackage[ps2pdf,%
a4paper=true,%
breaklinks=true,%
colorlinks=true,%
pdfauthor={First Author et al.},%
pdftitle={Template for manuscripts in Advances in Space Research}%
]{hyperref}

\begin{document}

\title{Neutrinos and cosmic rays}
\author{Thomas K. Gaisser \& Todor Stanev}
\address{Bartol Research Insitute and Dept. of Physics and Astronomy\\
University of Delaware, Newark, DE, USA}
\ead{gaisser@bartol.udel.edu}

\begin{abstract}
  In this paper we review the status of the search for high-energy
 neutrinos from outside the solar system and discuss the implications
 for the origin and propagation of cosmic rays.  Connections between
 neutrinos and gamma-rays are also discussed.

\end{abstract}

\maketitle

\section{Introduction}

Observation of high-energy neutrinos of astrophysical origin would
open a new window on origin of cosmic rays.  Neutrinos are expected at
some level in association with cosmic rays, both from interactions
of accelerated protons and nuclei in or near their sources and from interactions
of the cosmic rays during propagation in space.  Production of
 high-energy neutrinos requires interaction of hadrons to make mesons
 which then decay to neutrinos.  Observation of neutrinos from gamma-ray
 sources would therefore indicate a hadronic rather than electromagnetic
 origin for the photons.  Examples of possible sources are galactic
 supernova remnants and extragalactic objects such as gamma-ray bursts (GRB)
 and active galactic nuclei (AGN).

  In addition, wherever gamma-rays are produced
by interactions of cosmic rays during their propagation, neutrinos will also be
produced.  Examples of the latter are neutrinos related to the diffuse 
gamma-ray emission from the disk of the Milky Way~\cite{Stecker}
 and {\it cosmogenic} neutrinos
produced when cosmic rays of ultra-high energy (UHECR) interact with the
cosmic background radiation (CMB)~\cite{Berezinsky}. 
Both processes can be calculated
in a straightforward way.  For the Galaxy, the physics is pion production
in interactions of cosmic rays with gas in the interstellar medium,
 and the neutrino flux follows directly from the observed diffuse
 gamma-radiation from the same source.
The calculation of photo-pion production by protons in the cosmic microwave
 background (CMB) also follows from well-known physics, but in this case
 the level of neutrino production is highly uncertain because the ultra-high
 energy cosmic ray (UHECR) spectrum
 is unknown.  Whether there are sufficient protons above the threshold of
$3\times10^{19}$~eV is one of the main unanswered questions of neutrino
 astronomy.

The discovery of neutrino oscillations~\cite{SKnuosc} has important
implications for
neutrino astronomy.  One expects only muon and electron neutrinos to
be produced both in interactions with gas and in photo-pion production.
However, the effect of oscillations on an astronomical baseline is that the
initial flavor ratio evolves toward comparable numbers of all flavors
for the observer.  For example, for an initial flavor ratio of
$\nu_e:\nu_\mu:\nu_\tau\,=\,1:2:0$ the ratio at Earth would be
$1:1:1$~\cite{LearnedPakvasa}.  Since tau neutrinos are essentially absent
above $100$~GeV in the atmospheric
neutrino background, identification of a $\nu_\tau$ would be strong
evidence for astrophysical origin.  For this reason, the ability to
distinguish neutrino flavors is important.

\section{Status of searches for neutrino sources}

The biggest signal is expected in the muon neutrino channel.
Because of the long range of high energy muons, interactions
of $\nu_\mu$ outside the detector can produce muons that
reach and pass through the detector.  For an instrumented
volume even as large as 10~km$^3$, the external $\nu_\mu$
events are more numerous than interactions inside the instrumented
volume.  The most sensitive searches use the Earth as a filter
against the downward background of atmospheric muons by
requiring the muon track to be from below the horizon.

The most basic approach to neutrino astronomy is to look for
an excess of events from a particular direction in the sky.
AMANDA, Baikal, Antares and IceCube all make sky maps.  The search
can be binned or unbinned~\cite{Karle}.  After accounting
for the effective number of trials, no significant excess has
been seen in any detector.  A related approach is to look for
an excess of events from a list of objects selected because
they are likely neutrino sources.  The source list for IceCube~\cite{PtSrc},
for example, includes 13 galactic supernova remnants (SNR), and 30
extra-galactic objects, mostly AGN.  With its instrumented km$^3$
volume, IceCube is by far the most sensitive detector at present.
Published limits from IceCube during construction with 40
strings installed (IC-40) on specific point sources of neutrinos
in the Northern sky are less than $10^{-11}$cm$^{-2}$s$^{-1}$TeV$^{-1}$.
With the full IceCube the sensitivity is now approaching
 $10^{-12}$cm$^{-2}$s$^{-1}$TeV$^{-1}$, at which level TeV gamma-rays
are seen from some blazars such as Mrk 401~\cite{Mrk401}.

A related approach is to look for neutrinos correlated in time, either with 
each other or
with a gamma-ray event~\cite{Tdependent}.  The strongest limit from IceCube in
terms of constraining models that relate cosmic-ray origin with
production of neutrinos is the absence of neutrinos in coincidence
with GRB.  Recently data sets from two years of IceCube while the detector
was still under construction (IC-40 and IC-59) have been combined to obtain
a significant limit~\cite{I3GRB} on models~\cite{WB-GRB} in which GRBs are the main source of
extragalactic cosmic rays.  In total 215 GRBs reported
by the GRB Coordinated Network between April 5, 2008 and
May 31, 2010 in the Northern
sky were included in the search.  No neutrino was found during the
intervals of observed gamma-ray emission.

To set limits on the model~\cite{WB-GRB}, the expected neutrino spectrum
was calculated for each burst based on parameters derived~\cite{Guetta}
from features in the spectrum of the GRB.  In particular,
a break in the observed photon spectrum marks the onset of photo-pion
production by accelerated protons
interacting with intense radiation fields in the GRB jet.  
The neutrinos come from the decay of
charged pions.  Given a predicted neutrino spectrum, the expected
number of events was calculated for each burst.  The normalization
of the calculation is provided by the assumptions that a fraction
of the accelerated protons escape and provide the ultra-high energy
cosmic rays.  In the simplest case, the UHECR are injected
as neutrons from the same photo-production processes in which the
neutrinos are produced~\cite{Halzen1}.  With this normalization, 8 neutrinos are
expected in 215 GRBs and none is observed.

\begin{figure}[thb]
\begin{center}
\includegraphics[width=0.8\columnwidth]{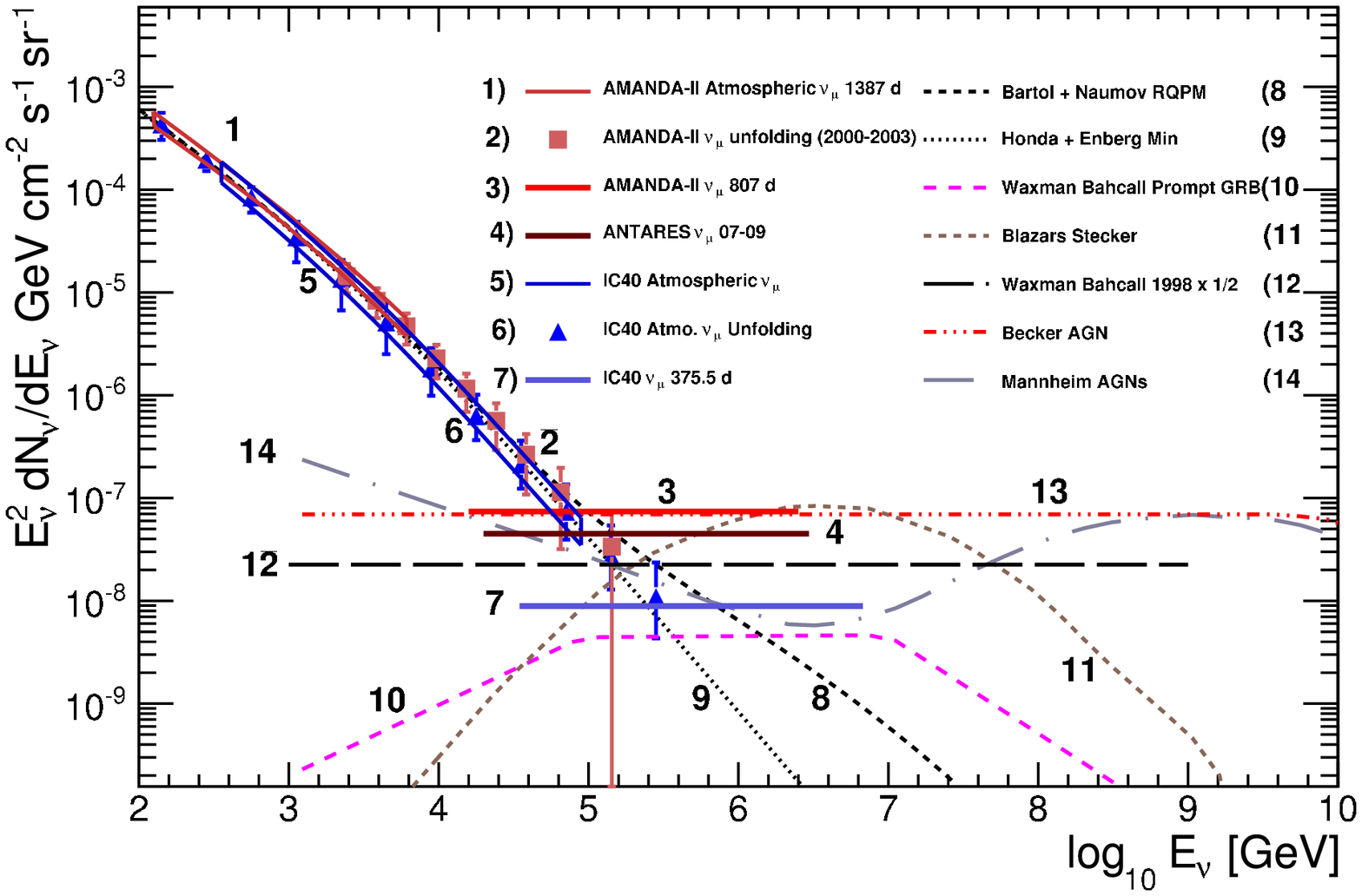}
\caption{Horizontal lines show limits on an $E^{-2}$ spectrum
of astrophysical muon neutrinos from AMANDA-II~\cite{AMANDA}, 
 Antares~\cite{Antares} and IceCube~\cite{IC40D}.  The plot
is from Ref.~\cite{IC40D} where full references are given.  
The limits are shown along with measurements
of the flux of atmospheric muon neutrinos and anti-neutrinos.  
}
\label{fig2}
\end{center}
\end{figure}

\section{Neutrinos from the whole sky}
It is important also to search for an excess
of astrophysical neutrinos from the whole sky at high energy above 
the steeply falling background of atmospheric neutrinos.  The
Universe is transparent to neutrinos, so the flux of neutrinos
from sources up to the Hubble radius may be large~\cite{Lipari}.
A toy model is helpful to illustrate this point.  Assume a distribution
of identical sources of neutrino luminosity $L_\nu$ (s$^{-1}$TeV$^{-1}$)
with a typical separation of order $d=10$~Mpc.  The flux from a
nearby source is $L_\nu/(4\pi d^2)$ (s$^{-1}$TeV$^{-1}$cm$^{-2}$). 
Integrating over the whole sky with a cutoff at the Hubble distance $D_H$
the flux from the whole sky is
\begin{equation} 
\phi\,\approx\,\int_{0}^{D_H}\,{\rho\,L_\nu \,r^2\over 4\pi r^2}{\rm d}\Omega{\rm d}r,
\label{diffuse}
\end{equation}
where $\rho \sim 1/d^3$ is the density of sources.  In this case the
ratio of the total flux of neutrinos from all directions to the flux from
a nearby source is $\sim 4\pi\,D_H\,/\,d\,\sim\,4000$ for $d=10$ Mpc.
Later we will cite examples of calculations for specific models of AGNs
and GRBs, which take account properly of red shift for distant sources.
In some cases the predicted diffuse fluxes are sufficiently high
to constrain the models more than the point source searches.
Before discussing the models, we first summarize the current status of the limits on diffuse
fluxes of high energy neutrinos.

The limit from IC-40, shown as the solid (blue) line \#7 in Fig.~\ref{fig2}, 
is from an analysis of approximately
14,000 upward neutrino-induced muons in IC-40~\cite{IC40D}.  This analysis 
proceeds by assuming a flux of neutrinos
with three components: conventional atmospheric neutrinos from decay of
kaons and pions; prompt neutrinos; and a hard spectrum of astrophysical
neutrinos assumed to have an $E^{-2}$ differential spectrum.  Free parameters
in fitting the data are the normalization of the prompt and astrophysical
neutrinos.  The normalization and slope of the atmospheric neutrinos are
also allowed to vary within a limited range.  The result is consistent with
conventional atmospheric neutrinos, with no need for a contribution from
prompt neutrinos and no evidence of a hard spectrum of astrophysical 
neutrinos.  A limitation of the analysis is that the atmospheric
neutrino background is represented by a simple power law extrapolation
of the calculation of Ref.~\cite{Honda} beyond $10$~TeV, and it averages
over all angles below the horizon.  

Also shown in Fig.~\ref{fig2} are several measurements of the 
flux of atmospheric neutrinos.  The fit for atmospheric neutrinos
from the IC-40 analysis that gives the diffuse limit is shown as a slightly
curved band extending from 0.33 to 84~TeV.  The reason that the diffuse
limit applies at much higher energy (39TeV to 7 PeV) is that it assumes
a hard, $E^{-2}$ differential energy spectrum for the neutrinos, in contrast
to the steep ($\sim E^{-3.7}$) atmospheric spectrum.
The other experimental
results on the high-energy flux of atmospheric $\nu_\mu+\bar{\nu}_\mu$
in Fig.~\ref{fig2} are from AMANDA~\cite{AMANDA1,AMANDA2} 
and IceCube-40~\cite{Warren}.  All the atmospheric neutrino spectra
shown here are averaged over angle.  The unfolding analysis of Ref.~\cite{Warren}
extends to $E_\nu\approx 400$~TeV.  The atmospheric fluxes shown are averaged
over the upward hemisphere.  At high energy atmospheric neutrinos from decay
of charged pions and kaons have a significant angular dependence (the ``secant theta"
effect) with the intensity increasing toward the horizon.
This angular dependence will be important in distinguishing atmospheric
background from astrophysical signal in future analyses.

At the current level of sensitivity
in the search for high-energy astrophysical neutrinos, the energy range where the 
atmospheric neutrino background becomes important is at 100~TeV and above,
as illustrated by the crossover of the limits and the atmospheric fluxes
in the Fig.~\ref{fig2}.
This energy is well beyond the range of detailed Monte Carlo
calculations~\cite{Honda,Bartol}, which extend only to $10$~TeV.
In addition, this is the energy range where prompt neutrinos from decay
of charm and heavier flavors may 
become important, but the expected level of this contribution is uncertain.
The spectrum of prompt neutrinos is harder by one power than the spectrum
of conventional atmospheric neutrinos in this energy range, and its angular
distribution is isotropic.  These features mimic a diffuse astrophysical
flux to some extent.  A possible strategy is to determine the level of
prompt lepton production with atmospheric muons which would remove the
ambiguity from this contribution to the background of atmospheric neutrinos.
Calculations that extend the atmospheric neutrino flux up to the PeV range
will also need to account for the primary composition in the knee region keeping
in mind that what is relevant is the spectrum of nucleons as a function of
energy per nucleon.

\begin{figure}[thb]
\begin{center}
\includegraphics[width=0.8\columnwidth]{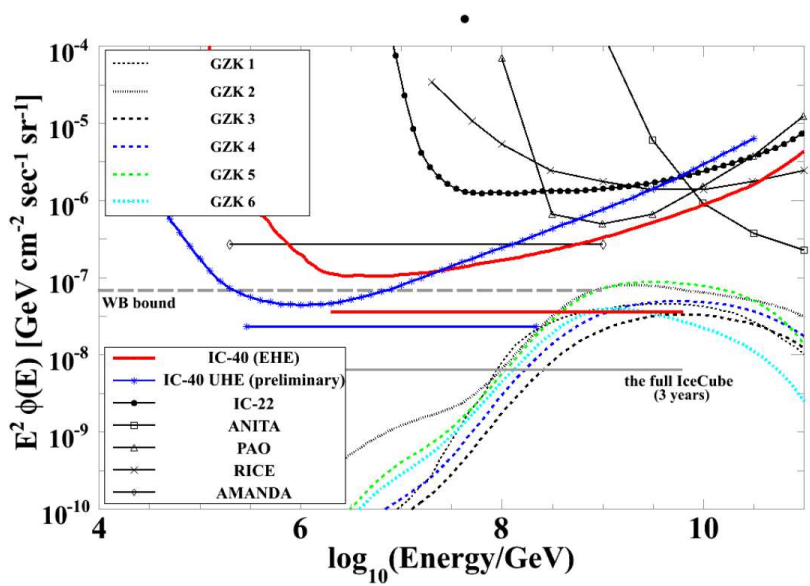}
\caption{ Collection of 
 limits on cosmogenic and ultra-high
energy neutrinos of all flavors.  The plot
is based on Ref.~\cite{EHE} where full references are given.
The extra curve included here, labeled {\em IC-40 UHE (preliminary)}
is from Ref.~\cite{Henrik}. 
}
\label{fig2a}
\end{center}
\end{figure}

Figure~\ref{fig2a} summarizes searches for neutrinos
of higher energy, including the region relevant for cosmogenic neutrinos.
Limits on the high energy side from Auger~\cite{Auger} 
and ANITA~\cite{ANITA,ANITAerratum} are shown as well
as the results at lower energy from IC-40.  In the IC-40 analysis
shown here~\cite{EHE}, the strategy is to look for extremely energetic events
where the atmospheric neutrino background should not be important.
The greatest sensitivity in this energy range is to events near the
horizon because vertically upward muons are absorbed by the Earth.
The contribution of $\nu_\tau$ is particularly important.  For the
Auger analysis, $\tau$ leptons produced by charged current interactions
of $\nu_\tau$ skimming the earth are expected to give the major
signal as the $\tau$ leptons decay over the array~\cite{AugerICRC}.  Around $10^6$~GeV
in IceCube an important contribution to an astrophysical signal
would come from $\nu_\tau$ regeneration in the Earth~\cite{HalzenSaltz}.
Simulations show that a significant fraction of the events generated
by cosmogenic neutrinos would appear as cascade-like events generated
by $\nu_e$ and $\nu_\tau$.  The appearance of $\nu_\tau$ depends on
energy.  For $E\approx 10^6$~GeV a charged current interaction inside IceCube
would give a ``double bang" event~\cite{LearnedPakvasa} with two
separated cascades, one when the $\nu_\tau$ interacts and the other
when the $\tau$ decays.  At lower energy there would be a single,
perhaps elongated, cascade and at higher energy a cascade plus the
track of a $\tau$-lepton either entering or leaving the instrumented volume.

The horizontal dashed lines in Figs.~\ref{fig2} and in Fig.~\ref{fig2a}
show a benchmark intensity, the {\em Waxman-Bahcall limit}.
Current IceCube limits are
below the original Waxman-Bahcall limit~\cite{WBlimit}.  We discuss the
implications of this fact later in the section on extragalactic sources.

\section{Production of astrophysical neutrinos}
 
 Since we have not yet detected neutrinos arriving to us from 
 astrophysical sources we have to use the existing gamma ray
 data to identify sources that are likely to produce neutrinos.
 There are two different ways to produce neutrinos in astrophysical
 sources. One is from interactions of accelerated protons and nuclei on matter.
 All kinds of mesons are produced and the charged mesons decay to
 muons and neutrinos while the neutral mesons decay mostly into 
 gamma rays. It is easy to do a rough estimate of the relation of 
 the neutrino and gamma ray fluxes from pion decay. If the gamma ray
 flux from $\pi^0$ decay is
 $\phi_\gamma \; = \; C \times E_\gamma^{-\alpha}$ the corresponding muon
 neutrino and antineutrino spectrum from $\pi^\pm$ decay is
 $\phi_\nu \; = \; C \times (1 - r_\pi)^{\alpha - 1} \times E_\nu^{-\alpha}$,
 where $r_\pi$ = $(m_\mu/m_\pi)^2$.  Since in astrophysical environments
 muons usually also decay, this
 flux is doubled and becomes roughly equal to the photon flux.  It is also
 straightforward to take into account the muons and neutrinos from decay
 of kaons.  The exact calculations are algebraically complicated because
 of polarization effects in muon decay~\cite{Lipari}.
 For a power-law distribution of protons with differential index $\alpha$
 the ratio of $\nu_\mu+\bar{\nu}_\mu$ to photons is $1.0$ for $\alpha = 2.0$ 
 and $0.7$ for $\alpha = 2.7$.
 We will start
 with the assumption that production via proton interactions
 in gas contributes the most to the
 neutrino production in many galactic sources (like supernova remnants
 and molecular clouds)
 where the matter density provides enough target for nuclear interactions.

 Neutrinos are also produced in interactions of protons 
 with ambient photons, $p \gamma \rightarrow N \pi$.  Possible photon
 backgrounds are those in jets of AGN and gamma ray bursts (GRB) as well as the
 CMB. The proton threshold energy for
 production of pions is
 $E_p^{thr} \; = \; {{m_\pi} \over {4 \epsilon}} (2 m_p + m_\pi)$, where
 $\epsilon$ is the energy of the photon in the lab system. 
 In the CMB ($\langle \epsilon \rangle$ = 6.3$\times$10$^{-4}$ eV) the
 proton  threshold 
 energy, calculated with the exact CMB spectrum is 3$\times$10$^{19}$ eV.
 It is more difficult to estimate the threshold energy and the secondary 
 particle spectra in AGN or GRB jets where the photon background spectra 
 usually have non-thermal, typically broken power-law spectra.
 One can simplify
 the estimate by assuming that all pion photoproduction goes through
 the $\Delta^+$, i.e.
 $p + \gamma \rightarrow \Delta^+ \rightarrow p \pi^0 (n p^+)$, which 
 is a reasonable approximation especially in the case of a steep
 proton spectrum interacting with thermal distribution of photons
 where most of the production occurs near the kinematic threshold.
 In the $\Delta^+$ approximation the production of neutral pions is
 twice that of $\pi^+$. The $\gamma$-rays 
 from $\pi^0 \rightarrow 2 \gamma$ decay would have higher energy than
 the neutrinos from the
 $\pi^+ \rightarrow \nu_\mu + \mu \rightarrow \bar{\nu}_\mu + \nu_e$ 
 decay chain.  In general, the ratio of the final energies of the
 $\gamma$-rays  to neutrinos is greater than one, but when production above
 the $\Delta$ resonance region is important,
 the ratio of charged to neutral pions is increased.
   
   An essential complication from the point of view of neutrino astronomy
   is that gamma-rays can also be produced in purely electromagnetic processes
   whenever accelerated electrons are present in the sources.  Synchrotron
   radiation is important at low energy and inverse Compton scattering at
   high energy, as well as bremsstrahlung when there is sufficient gas
   present to scatter the  electrons.

\begin{figure}[thb]
\begin{center}
\includegraphics[width=0.8\columnwidth]{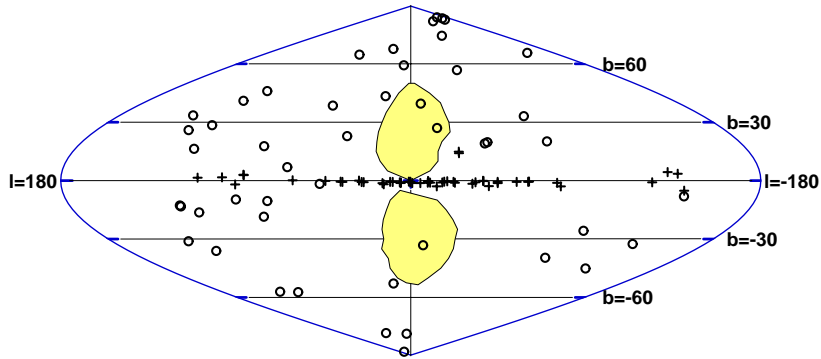}
\caption{Extragalactic (circles) and 
 galactic (pluses) TeV gamma ray sources from the TeVCat catalog.
 A large number of unidentified sources, some of which have 
 distance estimates, are not plotted. The two shaded regions
 indicate the Fermi bubbles.  There is also diffuse
 emission from the galactic plane, which is not shown here.
}
\label{TeVC}
\end{center}
\end{figure}

Figure~\ref{TeVC} shows the location of galactic and extragalactic sources
of TeV gamma rays from the TeVCat catalog~\cite{TeVCat}.
In the following sections we discuss examples of these objects as
potential neutrino sources.

\section{Galactic neutrinos}

\subsection{Neutrinos produced during propagation}
To the extent that the diffuse gamma radiation from the
plane of the Milky Way is due to interactions of 
cosmic rays with gas in the interstellar medium
through the $\pi^0$ channel, there will be a corresponding
level of diffuse neutrinos~\cite{Stecker}.  For power law
spectra there is a simple proportionality described in
the previous section whereby
\begin{equation}
{{\rm d}N_{\nu_\mu + \bar{\nu}_\mu}\over {\rm d}E_\nu}\,\sim\,
{{\rm d}N_\gamma\over{\rm d}E_\gamma}
\label{nu2gamma}
\end{equation}
 The newer analyses of the TeV $\gamma$-ray flux show that at 
 such high energy the production of $\gamma$-rays is dominated by 
 inverse Compton scattering of accelerated electrons

Recently the Fermi LAT collaboration has reported diffuse regions
of gamma emission extending to large distances both above
and below the galactic center as shown in Fig.~\ref{TeVC}.
There are competing models
of the gamma rays, one of which involves second order
Fermi acceleration of electrons by magnetic turbulence~\cite{Mertsch}
and the other which postulates a steady, long term production
of gamma rays by collisions of trapped cosmic rays with
diffuse gas~\cite{Crocker}.  In the latter case, there would
be a corresponding level of neutrino production given by Eq.~\ref{nu2gamma}.
If the energy spectrum of protons contained in the Fermi bubbles
is flatter than E$^{-2.3}$ they would produce neutrino fluxes
detectable in the Northern hemisphere~\cite{LunRaz}. The detection from 
the Southern hemisphere will be more difficult as a smaller portion
 of the bubbles is visible in upward going neutrinos.

 The Fermi/LAT collaboration studied the fraction of $\gamma$-rays 
 that are generated by protons during propagation by scattering
 on the galactic matter. They correlated this fraction of the
 gamma ray flux with
 the column density in a part of our Galaxy and established 
 an emissivity of 0.66$\times$10$^{-26}$ photons.s$^{-1}$sr$^{-1}$H-atom$^{-1}$
 above 300 MeV~\cite{LAT2}. Scaled to $\gamma$-rays of energy above 
 1 TeV with an E$^{-2.7}$ cosmic ray spectrum this gives an emissivity of 
 6.8$\times$10$^{-11}$ photons.s$^{-1}$sr$^{-1}$ for column density
 of 10$^{22}$ hydrogen atoms/cm$^{-2}$, very close to the old estimate~\cite{VSB1993}
 of 6$\times$10$^{-11}$. Since Ref.~\cite{VSB1993} has a model of the
 column density of the Galaxy we can now estimate the flux of neutrinos
 from the galactic plane taking account of the detector location. 

 IceCube can not see the inner galaxy in upgoing neutrinos. The closest
 it can get to it is $l$ = 31.8$^o$. We can define an area in longitude
 from 31.8$^o$ to 90$^{o}$ and latitude of 5$^o$ around the galactic plane
 that has an average column density of 1.8$\times$10$^{22}$ H-atoms/cm$^{-2}$.
 The angular area of that part of the galactic plane is 0.176 sr.
 The $\gamma$-ray flux above 1 TeV in this solid angle should be
 1.9$\times$10$^{-11}$ cm$^{-2}$s$^{-1}$ and the neutrino muon neutrino
 and antineutrino flux should be 0.7 of that for ($\alpha$ = 2.7, i.e.
 1.3$\times$10$^{-11}$ cm$^{-2}$s$^{-1}$). 
 
 Northern detectors, such as ANTARES or KM3Net, are able to observe
 the region of the inner galaxy. Cutting a similar region from
 $l$ = -90$^o$ to 90$^o$ we obtain an area of 0.546 sr with an
 average column density of 2.1$\times$10$^{22}$ Hydrogen atoms/cm$^{-2}$.
 The neutrino flux from that solid angle should be 
 4.8$\times$10$^{-11}$ cm$^{-2}$s$^{-1}$.
 
 To estimate the event rate we need to calculate the neutrino effective area
 of the detectors, which is defined so that, given a differential
 neutrino flux $\phi_\nu(E_\nu)$, the event rate is
 \begin{equation}
 {\cal R} = \int\,A_{\rm eff}(E_\nu)\times\phi_\nu(E_\nu)\,{\rm d}E_\nu.
 \label{Aeff}
 \end{equation}
 The effective area depends on neutrino flavor and accounts for the
 detector response as well as the physics of the neutrino propagation and
 interaction.  As an illustration here we consider $\nu_\mu+\bar{\nu}_\mu$ and
 an idealized detector that counts all muons above 100 GeV.  In this case
 \begin{equation}
 A_{\rm eff}(E_\nu)\,=\,P_\nu(E_\mu)\times\exp[-\sigma_\nu(E_\nu)N_AX(\theta)],
 \label{Aeff2}
 \end{equation}
 where $X(\theta)$ is the chord of the Earth at $\theta$ in g/cm$^2$, $N_A$ is
 the number of nucleons per gram and 
 \begin{equation}
 P_\nu(E_\mu)\,=\,N_A\,\int\,{\rm d}E_\mu{{\rm d}\sigma_\nu\over dE_\mu}(E_\mu,E_\nu)\,
 R_\mu(E_\mu)
 \label{Pnu}
 \end{equation}
 is the probability that a neutrino on a trajectory toward the detector produced
 a muon with enough energy to be detected and reconstructed.  The muon range is $R_\mu$.
 The $A_{\rm eff}$ we use is 0.5, 1.0, 55 m$^2$ at 1, 10 and 100 TeV respectively
 for an ideal detector with a projected physical area of one square kilometer.
 Assuming the diffuse galactic spectrum continues with a differential spectrum
 of $-2.7$ to $E\gg10$~TeV, the total flux from the region of the galactic plane
 visible to IceCube as estimated in the previous paragraph is $3\times 10^{-7}$~s$^{-1}$
 or ten events per year.  The flux is larger at Antares where the field of view
 includes the central region of the galaxy, but the projected area of the detector
 is much smaller than IceCube.  Assuming it is $\le 0.03$~km$^{2}$ there would
 be of order one event per year from this source in Antares.  Ninety per cent of
 the integral comes from $E_\nu<10$~TeV.

\subsection{Neutrinos from galactic cosmic-ray accelerators}

 Most of the galactic gamma ray sources shown in Fig.~\ref{TeVC}
 (23 of 52)
 are pulsar wind nebulae similar to the first detected TeV source,
 the Crab nebula. The $\gamma$-ray production in the Crab nebula
 has been modelled many times and the successful models are all
 purely electromagnetic. Because of that we do not expect 
 neutrino fluxes from such objects. There are also eleven 
 supernova remnants (SNR) with identified shell like morphology
 and seven sources identified as SNR/Molecular Clouds. All these
 sources are SNR with close by dense molecular clouds. It is not
 always obvious if the gamma ray production is in a part of the
 supernova remnant or only in the molecular cloud.

 The modeling of the gamma ray emission in supernova remnants 
 started in Ref.~\cite{Druryetal94}. It is based on the belief that 
 the cosmic ray energy spectrum at the source is flatter
 than what we observe at the Earth. As an example the calculation
 was applied to the Tycho (1572) shell like supernova remnant which is 
 likely to accelerate cosmic rays. Input parameters were the average
 SNR kinetic energy of 4.5$\times$10$^{50}$ ergs and the estimated 
 matter density of 0.7 cm$^{-3}$ around the remnant. If 20\% of all 
 accelerated cosmic rays interact around the supernova the 
 expected $\gamma$-ray flux above 1 TeV was estimated to be
 1.2$\times$10$^{-12}$ (E$_\gamma$/TeV)$^{-1.1}$ cm$^{-2}$s$^{-1}$.
 The Tycho supernova was detected much later. Its gamma ray flux
 has indeed a flat spectral index $\alpha$ = 1.95$\pm$0.5$\pm$0.3
 but much smaller flux at about 1\% that of the Crab nebula.
 (1 Crab unit~\cite{Aharetal2004} corresponds to integral flux of
 1.75$\times$10$^{-11}$ cm$^{-2}$s$^{-1}$ above 1 TeV.) 
 It is obvious that it either accelerates fewer 
 cosmic rays or contains them for a shorter time in the vicinity
 of the remnant. Defining a neutrino flux similar to such gamma ray flux 
 requires much bigger neutrino telescopes than we have.

 The HESS gamma ray collaboration published an analysis of their
 observation of the galactic center ridge~\cite{Aharetal2006} that
 partially explains why we have not seen as many $\gamma$-rays 
 from supernova remnants as were initially predicted. The HESS
 group determined that the $\gamma$-ray emission from that 
 part of the sky coincides with the positions of three molecular
 clouds with matter density of hundreds cm$^{-3}$. The total mass 
 in these clouds is 2-4$\times$10$^7$M$_\odot$. The observed
 $\gamma$-ray spectrum spectral index $\alpha$ = 2.3 is much flatter
 than the 2.7 spectrum we observe at Earth. The suggestion from
 this analysis is that we should look more at huge molecular clouds
 in the vicinity of supernova remnants rather than at the 
 supernova remnants themselves.
 
  In Ref.~\cite{GPS98}, the observed gamma-ray spectra from
 Egret~\cite{Egret} of  two supernova remnants associated with
 molecular clouds were modeled in detail by considering the
 contributions from bremsstrahlung and inverse Compton up-scattering by 
 electrons as well as photons from decay of neutral pions. 
 The SNR $\gamma$-Cygni, at a declination of +40$^o$ would give
 upward muons in IceCube.  It has recently been detected in TeV
 photons by VERITAS~\cite{VERITAS}.
 Here we use the fit corresponding to Fig.~5 of Ref.~\cite{GPS98},
 which assumed that $\gamma$-Cygni contains a single $\gamma$-ray source,
 to estimate the corresponding neutrino flux.  Although the model
 fit is not dominated by $\pi^0$ photons the fit predicts
 a flux of $2.5\times 10^{-12}$cm$^{-2}$s$^{-1}$ at 1 TeV with a
 cutoff above 10 TeV from this source.  The corresponding event
 rate, which is dominated by neutrinos in the range 0.3 to 3 TeV,
 is $3\times 10^{-8}$cm$^{-2}$s$^{-1}$, or one event per year, in
 IceCube.

\section{Neutrinos of extragalactic origin}


 An estimate of the maximum neutrino flux from extragalactic
 sources was made by Waxman \& Bahcall~\cite{WBlimit}.  
 High energy neutrinos come from interactions of higher energy nucleons.
 Therefore any source of high-energy neutrinos is a potential
 source of cosmic rays.  If the sources of the extra-galactic
 cosmic rays are transparent to nucleons so that they can
 inject cosmic rays into intergalactic space, then there is an implied limit on the associated
 flux of neutrinos from the condition that the sources not produce more cosmic
 rays than observed.  In models in which protons are contained in the sources
 by the magnetic fields essential to their acceleration, the limit is related
 to an estimate of the expected neutrino flux~\cite{TKG}.  Inside the accelerator,
 protons interact with photon backgrounds to photoproduce pions.  Secondary
 protons from $p+\gamma\rightarrow p + \pi^0$ may be trapped and
 reaccelerated in the jets, while secondary neutrons from $p+\gamma\rightarrow n + \pi^\pm$
 are not affected by the magnetic fields and may escape if the density of photons
 is not too high.  In such a situation there is a kinematic relation between
 the energy density in emitted neutrinos and the ultra-high energy cosmic rays
 from the decay of the neutrons.

Waxman \& Bahcall~\cite{WBlimit} used the observed spectrum of
UHECR to normalize their calculation.  Assuming an E$^{-2}$ spectrum,
 they estimated the power in cosmic rays in the energy range
 10$^{19}$ - 10$^{21}$ eV 
 as 5$\times$10$^{44}$ erg/year/Mpc$^3$.
 The upper bound of the neutrino 
 flux is calculated assuming that all accelerated protons have on average one
 photoproduction interaction in astrophysical jets. This leads to
 an upper limit of $\Phi_\nu E_\nu^2 = 1.5 \times 10^{-8}$
 GeV.cm$^{-2}$s$^{-1}$ster$^{-1}$.
 The upper bound increases by a factor of three if one assumes 
 cosmological evolution $(1 + z)^3$ for the sources of these ultrahigh
 energy cosmic rays (UHECR). There are different ways of looking at
 this calculation. The authors called it {\em an upper bound} but it can
 be in principle viewed as a lower limit because it only includes
 the neutrino production in photoproduction and additional neutrinos
 can be produced in $pp$ interactions and in UHECR proton interactions
 in the CMB. On the other hand, the original normalization was to
 the measurement by AGASA available at the time, which now appears
 to be an overestimate. The Hi-Res normalization would be a factor of 40\%
 lower and the Auger normalization a factor of two lower as measured 
 at 10$^{19}$ eV.

 The upper bound on the extragalactic neutrino flux was criticized 
 in Ref.~\cite{MPR} mostly because of the assumption of a flat E$^{-2}$
 injection spectrum for protons in the jets. The upper limit 
 derived in this paper has a more complicated shape that agrees
 with that derived in~\cite{WBlimit} only at E$_\nu$ = 10$^{18}$ eV. 
In any case, because of its simple form and normalization
to UHECR, the "bound" of Ref.~\cite{WBlimit} is a useful benchmark
for neutrino astronomy.  We return to this point after
discussing specific potential extragalactic sources of neutrinos. 

\subsection{Neutrinos from specific sources}

\subsubsection{Active Galactic Nuclei}

 Neutrino production in active galactic nuclei (AGN) is based on
 the assumption that the $\gamma$-ray fluxes detected from individual
 AGN are result of photoproduction interactions of protons that
 are accelerated in the AGN. The acceleration is often attributed to
 shock fronts in the jet that are generated by plasma blobs moving
 with different speeds. A different kind of model~\cite{Stecker2} 
 assumes that the acceleration of protons and their interactions happen
 at the shock created close to the central engine, where the 
 gravitationally attracted matter falling into the black hole meets
 the radiation pressure of the black hole emission. 
 The photoproduction interactions generate neutral pions
 that decay $\pi^0 \rightarrow 2 \gamma$ into 2 $\gamma$-rays
 and (mostly) $\pi^+$ whose decay chain generates $\nu_\mu$, $\bar{\nu}_\mu$,
 and $\nu_e$. 

 In such a hypothesis we have to look once again
 at the sources of TeV $\gamma$-rays
 and try to identify objects where these $\gamma$-rays are generated
 in proton photoproduction interactions in the jets, in the local
 photon fields or in $pp$ interactions in the environment of the
 object. Most of the extragalactic sources of TeV $\gamma$-rays shown in Fig.~\ref{TeVC}
 are blazers of different kinds. Blazars are active galactic nuclei (AGN)
 with jets pointing in our direction. Most of the $\gamma$-ray producing 
 blazars are high-frequency peaked BL Lac objects (HBL). The difference
 of these BL Lac objects with other blazars, such as the low-frequency 
 peaked LBL is the photo spectrum energy distribution. 

 Proton interactions in the jets of HBL and LBL objects were studied 
 in Ref.~\cite{AMetal03}. The theoretical calculations showed that 
 while both types of objects produce MeV to TeV $\gamma$-rays LBLs
 are favored for neutrino production. The main reason for that is the
 much higher photon density in LBLs compared to HBLs with similar luminosity.
 There are four LBLs on the map in Fig.~\ref{TeVC}: 
 APLib, S50716+714, 1ES1215+303, and the original BLLacertae. 
 The last three objects could be seen in upgoing neutrinos from the
 Southern hemisphere, although the neutrino fluxes from individual
 LBLs will be difficult to detect.
 The actual contribution to the diffuse neutrino fluxes depends 
 on the number of LBLs and HBLs in the Universe.
 
 Most fits to multi-wavelength spectra of AGN are made with electromagnetic
 processes only.  The low energy (radio - X-ray) part of the spectrum
 is explained as synchrotron radiation by electrons accelerated in the jets.
 The high energy (GeV-TeV) portion of the spectrum is fit with inverse
 Compton scattering by the same electron population boosting background
 photons to high energy.  The energy where the synchrotron component
 declines and the inverse Compton component becomes more important
 has been noted as a characteristic feature used to characterize
 different classes of AGN~\cite{Boetcher}.  In contrast, in the hadronic
 models of AGN discussed above, the high-energy portion of the
 gamma-ray spectrum is produced by a cascade initiated by
 the neutral pions produced by proton induced photoproduction.
 Mixed hadronic/electromagnetic models are also possible
 since it is likely that protons as well as electrons will be accelerated.
 
 An example where observations point to acceleration of protons is
 the ``orphan" flare of the AGN 1ES 1959+650~\cite{Boetcher}. 
 If the photon radiation
at all wavelengths is driven by the accelerated electrons, then when
a flare occurs both the synchrotron component and the inverse Compton
component should increase in unison, and this is often observed.
In this case there was a normal flare in both components and
later a sequence of flaring activity in the TeV component 
only observed by Whipple~\cite{Holder} and~\cite{Hegra}. 
 The Whipple group~\cite{Holder}  
reports an average flux of 0.64 in Crab units over a 60 day period, which
corresponds to a flux of gamma-rays with $E_\gamma\,>\,1$~TeV of 
$1.1\times 10^{-11}$cm$^{-2}$s$^{-1}$.  If we normalize the neutrino
spectrum to the gamma-ray spectrum measured at one TeV, we find
$\approx 2$ events would have been seen during this period in a kilometer
scale detector.  In this case, however, one cannot scale 
the expected neutrino flux to the gamma-rays in such a simple way, as noted in
the analysis of Ref.~\cite{HalzenOrphan}.  For one thing, the gamma rays
are likely to cascade in the intense electromagnetic radiation inside
the source.  In addition, the spectrum may be steepened by
interactions with extra-galactic background light between the source
and the Earth.  Reference~\cite{HalzenOrphan} addresses this problem
by assuming a canonical $E^{-2}$ spectrum for protons accelerated
in the source and hence for the neutrinos, which are not absorbed in the source.
They normalize the energy content of the neutrinos to the total energy
in the gamma-ray spectrum, which is quite steep.  The result also depends
on the Lorentz factor of the jet, so it is quite model dependent, but could
be much larger than the estimate from a one-to-one correspondence between
neutrinos and photons.

\subsubsection{Gamma-ray bursts}

 The processes that may generate neutrinos in gamma ray bursts (GRB)
 are not much different from these in AGN jets. The main differences
 include the much higher Lorentz factor of the GRB plasma (usually 
 set to an average value of $\Gamma$ = 300 compared to 10 in AGN jets),
 the short duration of the process (10 seconds), and the shape of the
 photon target spectrum known from the GRB detections. It is a broken
 power spectrum. The radiation below the break ($\epsilon_b$ = 1 MeV) follows
 a power law with index 1 and above the break it steepens to 2.
  This shape of the target photons generates a specific neutrino 
 spectrum. Protons of energy above $\Gamma^2 E_{thr}/\epsilon_b$
 ($E_{thr}$ is the proton interaction threshold in the co-moving
 frame)  interact
 mostly with the lower energy flat photon spectrum, while lower
 energy protons can only interact with the steeper energy spectrum
 higher energy photons.

 The resulting neutrino spectrum has two breaks, one at about 10$^5$ GeV
 where the neutrino energy spectrum changes from E$^{-1}$ to E$^{-2}$
 and another at about 10$^7$ GeV where the neutrino spectrum steepens
 because of the energy losses of the parent pions. 
 In the model of Ref.~\cite{WB-GRB} the middle part of the neutrino
 spectrum of all GRB (which are identical sources) is
 $\Phi_\nu E_\nu^2$ = 3$\times$10$^{-9}$ GeV.cm$^{-2}$s$^{-1}$ster$^{-1}$.

 The GRB studies are in rapid development both experimentally and 
 theoretically. A recent model of magnetized gamma ray bursts~\cite{GMesz}
 predicts detectable neutrino rates for GRBs containing significant 
 magnetic fields in their jets. The magnetic field contains the
 protons, which are reaccelerated, while the neutrons produced in
 photoproduction interactions leave the jets. Since their Lorentz factors
 start to differ, the protons and neutrons interact and generate 
 neutrinos on an almost E$^{-2}$ energy spectrum with an exponential
 cutoff at 250 GeV. Such neutrino spectra would produce
 of order one event
 in IceCube and Deep Core for GRBs at redshift of 0.1. The exact 
 flux magnitude and event rate depends strongly on the baryon load
 in the jet, the ratio of protons to electrons, and could be much
 lower.

\subsection{Implications of current limits}

In the discussion of potential Galactic sources of neutrinos, we gave
the example of $\gamma$-Cygni, a supernova remnant the environment
of which includes molecular clouds.  In the 
fit in which $\pi^0$ gamma-rays provide only part of the gamma flux,
we estimated only one neutrino per year in IceCube.
Quantitative estimates of neutrino fluxes from $\gamma$-ray
sources identified by
Milagro~\cite{Milagro} with $E_\gamma\sim$10~TeV
 lead to the prediction that IceCube should detect the corresponding
 neutrinos within three years~\cite{Halzenetal}.  A condition is that
 these sources accelerate cosmic rays to energies of 3 PeV/nucleon
 with a hard spectrum 
into the region of the knee
 of the spectrum.  Moreover, the photon flux observed in Milagro
 is assumed here to be entirely hadronic in origin.
 
 We also discussed the neutrino flux from production of pions in the disk of the galaxy
 during cosmic-ray propagation.  We estimate an excess of 10 neutrinos per
 year above the atmospheric background from a region that is 3\% of the Northern sky.
 This too will not be easy to detect as the full IceCube is expected
 to see more than a thousand atmospheric neutrinos per year from the same solid angle.

As noted above, one way to saturate the Waxman-Bahcall bound
is to have the protons trapped in
the acceleration region by the turbulent
magnetic fields needed to make the acceleration process work.
This scenario could be realized in jets
of GRB and of AGN if acceleration occurs in
internal shocks in the jets.  If this class of sources
is responsible for the UHECR, another implication would
be that the highest energy cosmic rays should be protons.
As IceCube limits become
increasingly strong, this class of models is constrained.

An example of a model already constrained by AMANDA,
the predecessor of IceCube, is that the nearby active
galaxy Cen A is typical of sources that contribute to
the extragalactic cosmic rays {\it and} that the cosmic rays are
accelerated inside the jets.  Several of the highest energy events
 observed by Auger come within a few degrees of Cen A~\cite{AugerCenA}.
Assuming that 2 out of 27 events with $E>57$~EeV
are accelerated in its jets, the corresponding neutrino
production is estimated in Ref.~\cite{Cuoco}.  Koers \& Tinyakov~\cite{Tinyakov}
follow through the consequences of this idea by assuming that
all UHECR come from sources like Cen A distributed throughout the Universe.
The argument schematically outlined in Eq.~\ref{diffuse} is used
to estimate the neutrino flux from all sources.  The source density
is normalized by requiring that the sum of all such sources give
the observed UHECR spectrum.  The predicted neutrino rate depends on
assumptions about cosmological evolution of the sources, but even
with no cosmological evolution, the level of neutrinos was comparable
to the AMANDA limits and is clearly ruled out by the current IceCube limits.

A generic alternative to acceleration of UHECR inside
the jets of AGN or GRB could be that they are
accelerated outside the jets, for example at the
termination shocks of AGNs.  In Ref.~\cite{Berezhko}, for example,
AGN are assumed to be the sources of extragalactic cosmic rays
with the acceleration occurring at the termination shocks
analogous to acceleration of galactic cosmic rays at SNR.
In this case the
composition of the extragalactic cosmic
radiation would depend on the ambient medium, and the level of
neutrino production would be contingent on the 
density of the surrounding medium and correspondingly low.

\subsection{Cosmogenic neutrinos}

\begin{figure}[thb]
\begin{center}
\includegraphics[width=0.8\columnwidth]{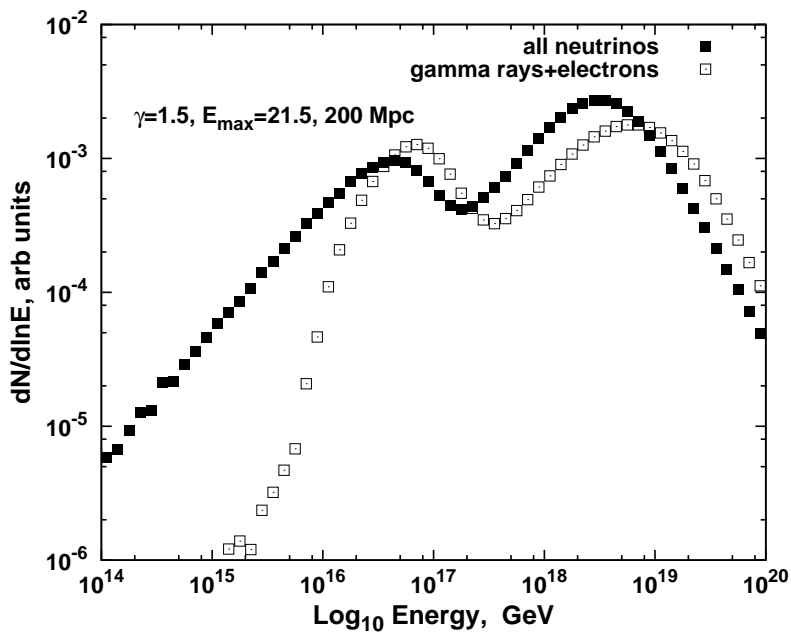}
\caption{ Neutrinos, gamma rays and electrons produced in the
 propagation of protons accelerated to a E$^{-2.5}$ spectrum 
 on 200 Mpc. 
}
\label{f325}
\end{center}
\end{figure}

 These ultrahigh energy neutrinos were suggested in 1969~\cite{Berezinsky}
 soon after the discovery of CMB. The UHECR interact in the microwave
 background in their propagation to us and produce pions and
 other mesons which later decay to neutrinos, electrons and gamma rays. 
 This source of neutrinos is independent of whether the
 UHECR are produced inside jets or at the termination
 shocks of AGN or GRB, or indeed from some other source altogether.
 The shape of the neutrino and $\gamma$-ray fluxes at production are shown in 
 Fig.~\ref{f325}.  We only show the fluxes of neutrinos, $\gamma$-rays
  and electrons produced in proton propagation on 200 Mpc ($z \simeq$= 0.05),
  a distance within which
the cascading process in CMB is completed. To obtain the total
  neutrino flux one has to account for the protons accelerated at earlier
  times and also to account for possible cosmological evolution on the
  proton accelerators. 
 The peaks of the spectra around 10$^{18}$-10$^{19}$ eV 
 are due to the muon neutrinos and antineutrinos generated in the
 meson and muon decays and to the neutral meson decays into $\gamma$-rays.
 The lower energy peaks just above 10$^{16}$ eV are due to $\bar{\nu}_e$
 and electrons from neutron decay. The peak $\gamma$-ray energy is 
 higher roughly by a factor of two than those of the
 $\nu_\mu$, $\bar{\nu}_\mu$ and $\nu_e$ 
 because the neutral pions decay in 2$\gamma$-rays while the charged
 pions decay to 3 neutrinos and one electron. 

 The exact flux of these {\em cosmogenic} neutrinos depends on many 
 factors, such as
 \begin{itemize}
\item The total emissivity of the Universe in UHE cosmic rays, usually
expressed in ergs/Mpc$^3$/year.
\item The average acceleration spectrum of these particles. The flatter
the spectrum is the more UHECR can interact in the CMB.
\item The maximum acceleration energy in the UHECR sources.
\item The cosmological evolution of the UHECR sources.
\item The chemical composition of UHECR.
 \end{itemize}
 If the highest energy cosmic rays are heavy nuclei, as
 suggested by the Auger Observatory measurements up to
 50 EeV, the energy spectrum of individual nucleons
 will cut off at relatively low energy which will
 decrease the fluxes of the $\geq$10$^{18}$ eV neutrinos.
 The flux of 10$^{16}$ eV $\bar{\nu}_e$ will increase because of the
 decay of neutrons from the spallation of the nuclei.
\begin{figure}[thb]
\begin{center}
\includegraphics[width=0.8\columnwidth]{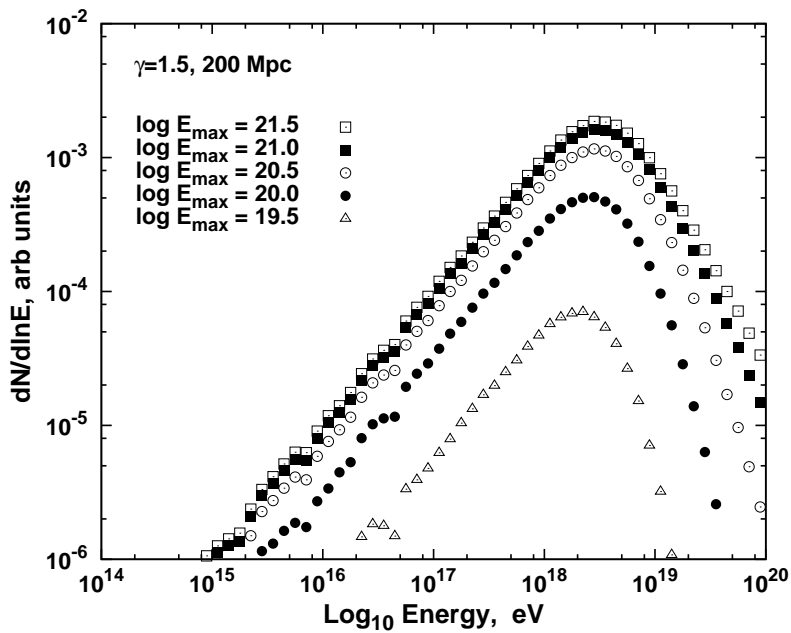}
\caption{ Muon neutrinos and antineutrinos generated in
 propagation of protons on 200 Mpc for different values 
 of the maximum proton energy at acceleration. 
}
\label{f326}
\end{center}
\end{figure}

 When calculated with the same input that Waxman \& Bahcall
 used~\cite{WBlimit} the flux of cosmogenic neutrinos touches
 the limit at the maximum of the muon neutrinos and antineutrinos
 and is generally lower at higher and lower energies~\cite{ESS}.
 The contemporary measurements of the UHECR spectrum show
 lower total emissivity and steeper acceleration spectrum
 than used in Ref.~\cite{WBlimit} which significantly decreases
 the expectations for cosmogenic neutrinos unless~\cite{SS05}
 the cosmological evolution of the UHECR sources is extremely strong.

Associated with the production of neutrinos during their propagation in the
cosmos is the predicted steepening of the spectrum 
 above $50$~EeV, often referred to as the {\em GZK effect}
 after the initials of the authors of the original 
 papers~\cite{GZK1,GZK2}.  A question that remains open
 is whether the observed steepening is the GZK effect (as often
assumed) or whether it is simply the sources of UHECR reaching
their maximum energy.  In this connection it is relevant to
recall the Hillas diagram~\cite{Hillas84} which illustrates
the difficulty of accelerating particles to $100$~EeV.
As an illustration of the importance of this point for neutrino
astronomy, we show in
Fig.~\ref{f326} how the predicted
fluxes of cosmogenic $\nu_\mu+\bar{\nu}_\mu$ depend on maximum energy 
assumed at the sources.
Muon neutrinos and
 antineutrinos are generated in proton propagation over 200 Mpc. 
 The spectral index $\alpha$ = 2.5 and there is an exponential 
 decline of the flux at different values of $\log_{10} E_{max}$
 from 21.5 to 19.5~eV/nucleon. There are two effects: the maximum of the 
 neutrino flux moves to lower energy when E$_{max}$ decreases, and
 the total flux of cosmogenic neutrinos also decreases. While the 
 maximum neutrino flux for $\log_{10}E_{max}$ = 21.5 is at 2$\times$10$^{-3}$
 it decreases to 5$\times$10$^{-5}$ for  $\log_{10}E_{max}$ = 19.5.
 There are no cosmogenic muon neutrinos and antineutrinos
 generated in the local Universe if $\log_{10}E_{max}$ = 19.
 
 \section{Conclusion}
 The full IceCube detector has been operating since May, 2010.
 This means that the integrated exposure under analysis will
 increase quickly compared to approximately one km$^3$ year
 from IC-40 and IC-59 currently under analysis.  On the horizon
 are plans for KM3NeT~\cite{KM3NeT} in the Mediterranean and GVD in Russia~\cite{GVD}
 that would provide kilometer-scale coverage from Northern mid-latitudes.
 Installation of the Askaryan Radio Array next to IceCube started
 recently~\cite{ARA}.  The goal is to achieve sensitivity corresponding
 to $200$~km$^2$ area sensitive to cosmogenic (GZK) neutrinos.
 
 An important modification of the original plan of IceCube was
 the installation of 8 specially instrumented strings 
 with their optical modules concentrated in the deep, clearest
 ice in between the 7 standard strings in the center of IceCube.
 Together these 15 strings form the DeepCore of IceCube~\cite{DeepCore}.
 The full year of DeepCore data in 2010-2011 with 73 standard strings
 and 6 special stings in place has already been analyzed~\cite{Ha}.
 By using the surrounding detectors of IceCube to veto atmospheric muons,
 it has been possible for the first time to identify neutrino-induced 
 cascade event in IceCube.  These are interactions with a mean energy
 of $180$~GeV that include charged current interactions of electron
 neutrinos and neutral current interactions of all flavors.
 Simulations show that about $40$~\% of the sample consists of
 charge-current interactions of $\nu_\mu$ inside the DeepCore region.
 The goal is to measure neutrinos with energies between $10$~GeV
 and a TeV.  This would allow studies of neutrino oscillations
 and improved sensitivity for soft neutrino sources, including the
 Southern sky.  Proposals for installing still more densely spaced
 detectors to lower the threshold further are being discussed~\cite{Ty,Darren}.
 
 From the point of view of neutrino astronomy, the obvious goals are
 \begin{itemize} 
 \item to detect neutrinos from sources in the Milky Way in support
 of the quest to understand the origin of galactic cosmic rays;
 \item to detect neutrinos from GRB and/or AGN \\ --or to make
 the limits sufficiently low compared to observed gamma-ray fluxes
 to rule out acceleration inside relativistic jets as the primary
 source of UHECR; and
 \item to detect cosmogenic neutrinos\\ --or to make
 the limits sufficiently low to constrain the upper limit
 of the energy per nucleon of UHECR.
 \end{itemize}


\begin{thebibliography}{99}
\bibitem{Stecker} F.W. Stecker, Astrophys. J. {\bf 228} 919 (1979).
\bibitem{Berezinsky} V.S. Berezinsky \& G.T. Zatsepin, Phys. Lett. {\bf 28} B, 423 (1969).
\bibitem{SKnuosc} Y. Fukuda et al. (Superkamiokande Collaboration) Phys. Rev. Letters
{\bf 81} 1562 (1998).
\bibitem{LearnedPakvasa} J.G. Learned \& S. Pakvasa, Astropart. Phys. {\bf 3}, 267 (1995).
\bibitem{Karle} J. Braun et al., Astropart. Phys. {bf 29}, 299 (2008).
\bibitem{PtSrc} R. Abbasi et al., (IceCube Collaboration) Astrophys. J. {\bf 732}, 18 (2011).
\bibitem{Mrk401} M.K, Daniel (for the Veritas Collaboration) Proc. of the 4th
Heidelberg International
Symposium on High Energy gamma-Ray Astronomy 2008 (arXiv:0810.0387).
\bibitem{Tdependent} R. Abbasi et al. (IceCube Collaboration) Astrophys. J. {\bf 744}, 1
(2011).
\bibitem{I3GRB} A. Franckowiak (for the IceCube Collaboration), Proc. 2011 Fermi Symposium,
eConf C110509 (arXiv:1111.0335).
\bibitem{WB-GRB} E. Waxman \& J.N. Bahcall, Phys. Rev. Letters {\bf 78}, 2292 (1997).
\bibitem{Guetta} D. Guetta et al., Astropart. Phys. {\bf 20}, 429 (2004).
\bibitem{Halzen1}  M. Ahlers, M.C. Gonzalez-Garcia \& F. Halzen, 
Astropart. Phys. {\bf 35}, 87 (2011).
 \bibitem{AMANDA} R. Abbasi et al. (IceCube Collaboration) Phys. Rev. D {\bf 79}, 062001 (2009).
 \bibitem{Antares} M. Ageron et al (Antares Collaboration) arXiv:1104.1607v2.

\bibitem{IC40D} R. Abbasi et al. (IceCube Collaboration),
Phys. Rev. D {\bf 84}, 082001 (2011).
\bibitem{AMANDA1} R. Abbasi et al. (IceCube Collaboration), 
Phys. Rev. D {\bf 79}, 102005 (2009).
\bibitem{AMANDA2} R. Abbasi et al. (IceCube Collaboration), 
Astropart. Phys. {\bf 34}, 48 (2010).
\bibitem{Warren} R. Abbasi et al. (IceCube Collaboration), 
Phys. Rev. D {\bf 83}, 012001 (2011).

\bibitem{Honda} M. Honda et al., Phys. Rev. D {\bf 75}, 043006 (2007).
\bibitem{Bartol} G.D. Barr et al., Phys. Rev. D {\bf 70}, 023006 (2004).
\bibitem{EHE} R. Abbasi et al. (IceCube Collaboration), 
Phys. Rev. D {\bf 83}, 092003 (2011).
\bibitem{Henrik} {\em Searching for an Ultra High-Energy Diffuse Flux of 
Extraterrestrial Neutrinos with IceCube 40},
H. Johansson, Ph.D. Thesis, University of Stockholm (2011).

\bibitem{Auger} The Pierre Auger Collaboration, Phys. Rev. D {\bf 79}, 102001
 (2009). Proc. 32nd Int. Cosmic Ray Conf. (Beijing, 2011) arXiv:1107.4805.
\bibitem{ANITA} P. Gorham et al., Phys. Rev. D {\bf 82}, 022004 (2010).
\bibitem{ANITAerratum} P. Gorham et al., arXiv:1011.5004.
\bibitem{AugerICRC} Y. Guardincerri (for the Pierre Auger Collaboration),
\bibitem{HalzenSaltz} F. Halzen \& D. Saltzberg, Phys. Rev. Letters {\bf 81}, 4305 (1998).
\bibitem{WBlimit} E.~Waxman \& J.N. Bahcall, Phys. Rev. D{\bf 59} 023002 (1999).
\bibitem{TKG} T.K. Gaisser, arXiv:astro-ph/9797283m 25 July, 1997.

\bibitem{Lipari} P. Lipari, Astropart. Phys. {\bf 1}, 159 (1993). 
\bibitem{TeVCat} http://tevcat.uchicago.edu
\bibitem{Mertsch} P. Mertsch \& S. Sarkar, Phys. Rev. Lett. {\bf 107}, 091101 (2011).
\bibitem{Crocker} R.M.~Crocker \& F.~Aharonian, Phys. Rev. Lett., {\bf 106}
 101102 (2011)
\bibitem{LunRaz} C.~Lunardini \& S.~Razzaque, arXiv:1112.4799
\bibitem{LAT2} A.A.~Abdo et al (Fermi/Lat Collaboration), Ap.J. {\bf 703},
 1249 (2009)
\bibitem{VSB1993} V.S.~Berezinsky et al, Astropart. Phys. {\bf 1}, 281 (1993)
\bibitem{Druryetal94} L.O'C. Drury, F.A. Aharonian \& H. V\"{o}lk,
 Astron. Astrophys. {\bf 289}, 959 (1994)
\bibitem{Aharetal2004} F.~Aharonian et al. (HESS Collaboration),
 Ap. J. {\bf 614} 807 (2004)
\bibitem{Aharetal2006} F.~Aharonian et al (HESS Collaboration)
\bibitem{GPS98} T.K. Gaisser, R.J. Protheroe \& Todor Stanev, Astrophys. J. {\bf 492}, 219 (1998).
\bibitem{Egret} J.~Esposito et al., Astrophys. J. {\bf 461}, 820 (1996).
\bibitem{VERITAS} A. Weinstein (for the VERITAS Collaboration) Proc. 32nd ICRC (Beijing),
arXiv:1111.2093
\bibitem{MPR} K.~Mannheim, R.J.~Protheroe \& J.P.~Rachen, Phys. Rev. D
\bibitem{Stecker2} F.W.~Stecker et al, Phys. Rev. Lett {\bf 66}, 2697 (1991;
 err.-ibid {\bf 69}, 2738 (1992)
\bibitem{AMetal03} A.~M\"{u}cke et al, Astropart. Phys. {\bf 18}, 593 (2003)
\bibitem{Boetcher} M. B\"{o}tcher, Astrophys Space Sci {\bf 309}. 95 (2007)
\bibitem{Holder} J. Holder et al., Astrophys. J. {\bf 583}, L9 (2003)
\bibitem{Hegra} H. Krawczynski et al., Astrophys. J. {\bf 601}, 151 (2004). 
Astron. Astrophys {\bf 406}, L9 (2003)
\bibitem{HalzenOrphan} F. Halzen \& D. Hooper, Astropart. Phys. {\bf 23}, 537 (2005).
\bibitem{Milagro} A.A. Abdo et al., Astrophys. J. {\bf 658}, L33 (2007) 
\bibitem{Halzenetal} M.C. Gonzalez-Garcia, F. Halzen \& S. Mohapatra,
Astropart. Phys. {\bf 31}, 437 (2009).
\bibitem{AugerCenA} P.~Abreu et al. (Auger Collaboration), Astropart. Phys.
 {\bf 34}, 314 (2010)
\bibitem{GMesz} S.~Gao \& P.~Meszaros, arXiv:1112.5664
\bibitem{Cuoco} A.~Cuoco \& S.~Hannestad, Phys. Rev. D{\bf 77}:123518 (2008)
\bibitem{Tinyakov} H.B.J.~Koers \& P.~Tinyakov, Phys. Rev. D{\bf 78}:083009
 (2008)
\bibitem{Berezhko} E.G. Berezhko, Astrophys. J. {\bf 698}, L138 (2009)
\bibitem{ESS} R.~Engel, D.~Seckel \& T.~Stanev, Phys. Rev. D{\bf 64}:093010
 (2001)
\bibitem{SS05} D.~Seckel \& T.~Stanev, Phys. Rev. Lett. {\bf 95}:141101
 (2005)
\bibitem{GZK1} K.~Greisen, Phys. Rev. Lett. {\bf 16}, 748 (1966)
\bibitem{GZK2} G.T.~Zatsepin \& V.A.~Kuz'min, JETP Lett. {\bf 4}, 78 (1966)
\bibitem{Hillas84} A.M.~Hillas, Ann.Rev.Astron.Astrophys. {\bf 22}, 425 (1984)
\bibitem{KM3NeT} U. Katz, Nucl. Inst. Meth. A {\bf 567}, 457 (2006). 
 See also KM3NeT Technical Design Report
 at http://www.km3net.org/TDR/
\bibitem{GVD} V. Ayutdinov (for the Baikal Collaboration) arXiv:0811.1110
\bibitem{ARA} P. Allison et al., arXiv:1105.2854
\bibitem{DeepCore} IceCube Collaboration, arXiv:1109.6096
\bibitem{Ha} C. Ha (for the IceCube Collaboration) arXiv:1201.0801
\bibitem{Ty} T. DeYoung (for the IceCube Collaboration) arXiv:1112.1053
\bibitem{Darren} D. Grant (for the IceCube Collaboration) in arXiv:1111.2742
\end{thebibliography}
\end{document}